\documentclass[twocolumn,nofootinbib,amsmath,prd,aps,superscriptaddress,tightenlines,preprintnumbers]{revtex4}
\usepackage{graphicx}
\usepackage{epstopdf}
\usepackage{amsfonts}
\usepackage{bm}
\usepackage{epsfig}
\usepackage{graphics}
\usepackage{xspace}
\usepackage[usenames]{color}
\newcommand{\roughly}[1]{\mathrel{\raise.3ex\hbox{$#1$\kern-0.85em
\lower1ex\hbox{$\sim$}}}}

\newcommand{\lsim}{\roughly<}

\def\SM{{\scriptscriptstyle SM}}

\def\ba{\begin{eqnarray}}
\def\ea{\end{eqnarray}}
\def\be{\begin{equation}}
\def\ee{\end{equation}}

\def\ssH{{\scriptscriptstyle H}}
\def\ssI{{\scriptscriptstyle I}}
\def\ssL{{\scriptscriptstyle L}}

\def\ssT{{\scriptscriptstyle T}}

\def\ssZ{{\scriptscriptstyle Z}}

\def\A{\mathcal{A}}

\def\G{\mathcal{G}}
\def\H{\mathcal{H}}

\def\R{\mathcal{R}}

\def\nn{\nonumber}

\def\({\left(}
\def\){\right)}
\def\eps{\epsilon}

\def\pref#1{(\ref{#1})}

\begin{document}

\newcount\hour \newcount\minute
\hour=\time \divide \hour by 60
\minute=\time
\count99=\hour \multiply \count99 by -60 \advance \minute by \count99
\newcommand{\mydate}{\ \today \ - \number\hour :00}

\preprint{CERN - PH - TH / 2010 - 033}
\preprint{Pi-Partphys-174}

\title{On Higgs Inflation and Naturalness.}
\author{C.P. Burgess,${}^{1,2}$ H.M. Lee${}^3$ and
Michael Trott${}^1$ \\
${}^1$ Perimeter Institute for Theoretical Physics, Waterloo ON,
N2L 2Y5, Canada.\\
${}^2$ Dept. of Physics \& Astronomy, McMaster University,
Hamilton ON, L8S 4M1, Canada.\\
${}^3$ CERN, Gen\`eve 23, CH-1211, Switzerland. }

\date{\today}

\begin{abstract}
We  reexamine recent claims that Einstein-frame scattering in the Higgs inflation model is unitary above the cut-off energy $\Lambda \simeq M_p/\xi$. We show explicitly how unitarity problems arise in both the Einstein and Jordan frames of the theory. In a covariant gauge they arise from non-minimal Higgs self-couplings, which cannot be removed by field
redefinitions because the target space is not flat. In unitary gauge, where there is only a single scalar which {\em can} be redefined to achieve canonical kinetic terms, the unitarity problems arise through non-minimal Higgs-gauge couplings.
\end{abstract}
\maketitle

\bigskip

\section{Introduction}

The idea that non-minimal coupling of a scalar to gravity might lead to successful Inflation \cite{earlierNMC} has been reexamined recently with the very economical proposal of applying it to the Standard Model (SM) Higgs boson ($\H$) \cite{Hinfa,HinfRG}. In this scenario, the addition of the Higgs-gravity interaction, $\delta \mathcal{L} = -\xi (\H^\dagger \H) \, R$, to the Einstein-Hilbert
and Standard Model Lagrangians is used to obtain an inflationary slow roll. The additional parameter $\xi \simeq 10^4$ can be adjusted to align the spectrum of primordial perturbations with WMAP constraints \cite{WMAP}.

It is natural to wonder whether a coupling as large as $\xi \simeq 10^4$ could jeopardize the validity of the classical
approximation, on which the claim of inflationary behaviour is based. In ref.~\cite{Burgess:2009ea} it was shown that standard power-counting techniques \cite{GREFT} imply that semiclassical perturbation theory must break down at energies at or below the scale $\Lambda \simeq M_p/\xi$ (where $M_p = 2.44 \times 10^{18} \, {\rm GeV}$ is the reduced Planck mass), in agreement with explicit calculations \cite{HW}. Semiclassical scattering amplitudes violate unitarity above this energy. This scale can be regarded as an upper bound on the energy domain over which the theory can be regarded as a weakly coupled effective field theory.

Above this scale something new must intervene (such as new degrees of freedom, or perhaps strong coupling \cite{Bezrukov:2009db}), but whatever it is it must successfully compete with classical effects derived in the low-energy theory in order to solve the basic unitarity problem. Unfortunately, the Hubble scale during
the putative inflationary regime is also of order $H_\ssI \simeq \Lambda$ \cite{Hinfa}, thus non-adiabatic effects due to the universal expansion can be competitive with the effective theory's cutoff, putting into question the validity of semi-classical calculations within the effective theory (on which the inference of inflationary behaviour is based). Of course it might be that whatever the new high-energy physics is, it fixes the unitarity problem without ruining the inflationary conclusions. But whether this is possible remains to be shown.

Ref.~\cite{Barbon:2009ya} raised a related objection to Higgs inflation proposals by performing the explicit field redefinitions required to canonically normalize the metric and scalar degrees of freedom, identifying how the scale $\Lambda$ explicitly appears in the resulting scalar potential. In particular they argued that $\Lambda$ controls the expansion of the potential in powers of the physical fields, in addition to controlling the low-energy
(derivative) expansion itself. Since inflation occurs at field values that are large compared with $\Lambda$, this led to the conclusion that the shape of the potential is inadequately known in the required regime to ensure a slow roll occurs without fine tuning.

This argument that the scale $\Lambda$ controls the small-field expansion of the scalar potential is closely related to our unitarity argument, which argues that $\Lambda$ controls the low-energy approximation. In general the small-field expansion need not be controlled by the same scale as is the low-energy, derivative, expansion. Supersymmetric theories with flat directions provide the simplest examples where these scales differ, and when they do it is usually because the scalar potential is very shallow; large field values do not cost much energy. However, it is precisely because the Higgs potential is {\em not} particularly shallow these two scales are usually fairly
close to one another in Higgs physics \cite{BE}.

Recently Ref.~\cite{Lerner:2009na} has challenged these claims, arguing that there are no unitarity problems at energies $E \simeq \Lambda$ in the Einstein frame.\footnote{The Jordan frame is the
defining frame of the theory with a non-canonical graviton, while the Einstein frame is defined by performing a Weyl rescaling of the metric so that the graviton field is canonically normalized.}
They rightly point out that the explicit scattering calculations of ref.~\cite{HW} were performed in the Jordan frame, and argued that because physical observables cannot depend on how we define our fields, the appearance of unitarity violations must be present (or absent) in all frames. They argue that the absence of a problem in the Einstein frame means that there might be more to unitarity violation in the Jordan frame than meets the eye.

In this note we briefly revisit these issues, motivated by the attention that Ref.~\cite{Lerner:2009na} has received. Although our analysis in \cite{Burgess:2009ea} was performed generally enough to apply to either frame, we here explicitly identify the leading contributions to scalar scattering amplitudes in the Einstein frame, to see if these reproduce the energy dependence found by the Jordan-frame calculation of \cite{HW}. Contrary to the claims of \cite{Lerner:2009na}, we find that they do, although their origin is somewhat subtle. In particular, they would be missed by a naive analysis of the Higgs potential in terms of the single, real field that describes the physical Higgs, ignoring the existence of the remaining degrees of freedom of the Higgs doublet.

\section{Unitarity calculations}

Recall that the theory of interest is defined by the action
\be
 \frac{ \mathcal{L}_{H\,{\rm inf}}}{\sqrt{- \hat g} }
 = \mathcal{L}_{\scriptscriptstyle SM} - \left[ \frac{M_p^2 }{2}  +
 \xi \, (\H^\dagger \H ) \right] \hat R \,,
\ee
where $\xi$ is a dimensionless coupling, $\H$ is the usual Standard Model Higgs doublet and $\hat g_{\mu\nu}$ is the Jordan-frame metric.\footnote{Our signature is mostly plus, and we use Weinberg's curvature conventions.}

Only the Higgs sector is required of the Standard Model
lagrangian, including the quartic Higgs potential,
\be
 \mathcal{L}_\SM = - \hat g^{\mu \, \nu}
 \, (D_\mu \H)^\dagger \, (D_\nu \H)
 - \lambda_\ssH \left( \H^\dagger \H
 - \frac{v_\ssH^2}{2} \right)^2,
\ee
where $D_\mu$ denotes the usual covariant derivative and
$\lambda_\ssH$ is related to the Higgs boson mass by $m_\ssH^2 \simeq 2 \lambda_\ssH v_\ssH^2$. In the Higgs inflationary literature it is standard to specialize immediately to unitary gauge, for which the Higgs doublet is given by $\sqrt2 \, \H = (0, h)^{\scriptscriptstyle T}$ with $\langle h \rangle = v_\ssH$, but when working in the Einstein frame we find it instructive not to do so until later in the calculation.

\begin{figure}[hbtp]
\centerline{\scalebox{0.75}{\includegraphics{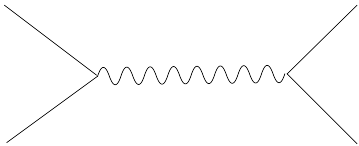}}}
\caption{Scalar scattering at high energies that leads to
unitarity violation at the scale $\sim \Lambda$ in the Jordan frame.}
\end{figure}

\subsection{Jordan Frame}

We first sketch the form of the Jordan-frame result, for which it is useful to choose unitary gauge. At large energies the dominant tree-level contribution comes from the graph of Fig.~1, with the external lines denoting physical Higgs particles and the internal line representing a graviton propagator. The vertex is obtained by directly expanding the $\xi (\H^\dagger \H) \hat R$ term about Minkowski space, using $\hat g_{\mu \, \nu} = \eta_{\mu\nu} + \kappa \, h_{\mu \, \nu}$ with $\kappa \propto 1/M_p$, to get 
\ba
 \delta  \mathcal{L} &=& - \sqrt{- \hat g} \; \xi (\H^\dagger \H) \,\hat R, \nn \\
 &=& \frac{\xi}{M_p} \, h^2 \, \eta^{\mu\nu}
 \, \partial^2 \, h_{\mu \nu} \, \, + \cdots ,
\ea
showing its dependence on $\Lambda = M_p/\xi$. Evaluation of this graph reproduces the high-energy part of the result of \cite{HW}, giving a scattering amplitude of the form
\be
 \A (E) \simeq (E^2/\Lambda)^2 (1/E^2) \simeq (E/\Lambda)^2 \,,
\ee
which gives a cross section $\sigma(E) \simeq E^2/\Lambda^4$. This remains below the unitarity bound provided $E \lsim \Lambda$, in agreement with \cite{HW} and the power-counting analysis of \cite{Burgess:2009ea}.\footnote{A potential loophole to this argument would be cancelation of the most singular energy growth amongst different graphs. Although such cancelations arise for $hh \rightarrow hh$ scattering at tree level for a single scalar field, $h$, \cite{Huggins:1987ea} it does not occur for all of the
components of the Higgs doublet. See the addendum below for more discussion of this issue.}

\subsection{Einstein Frame}

We now repeat the analysis in the Einstein frame, which is
obtained by using the Weyl transformation $\hat g_{\mu\nu} \to g_{\mu\nu}$, where
\be
 \hat g_{\mu\nu} = f \, g_{\mu\nu} \,,
\ee
with
\be
 f = \left[1 + 2 \, \xi ( \H^\dagger \H) /M_p^2 \right]^{-1} \,.
\ee
A short calculation shows that after such a transformation the Lagrangian becomes (after dropping total derivatives)
\ba
 \frac{ \mathcal{L}_{H\,{\rm inf}}}{\sqrt{-g}} &=& - \frac{1}{2}
 \, M_p^2 \, R  - \lambda_\ssH f^2 \left( \H^\dagger \H
 - \frac{v_\ssH^2}{2}  \right)^2 \,, \\
 &\,& \hspace{-2cm} - g^{\mu \nu}
 \left[ f (D_\mu \H)^\dagger \, (D_\nu \H)
  + \frac{3 \, \xi^2 f^2}{2 \, M_p^2} \,
 \partial_\mu (\H^\dagger \, \H) \,
 \partial_\nu(\H^\dagger \, \H) \right].\nn
\ea
Expanding the last line of this expression in the small field limit, $\xi \, (\H^\dagger \H)/M_p^2 \ll 1$, gives the following dimension-six interactions that are of later interest
\ba \label{dim6}
 \frac{\Delta  \mathcal{L}_{H\,{\rm inf}}}{\sqrt{-g}} &=& \frac{2 \, \xi}{M_p^2}
 \, g^{\mu \nu} \,
 (D_\mu \H)^\dagger \, (D_\nu \H) \,
 (\H^\dagger \H) \\
 && \qquad - \frac{3 \, \xi^2 }{2 \, M_p^2} \, g^{\mu\nu} \,
 \partial_\mu (\H^\dagger \, \H) \,
 \partial_\nu(\H^\dagger \, \H). \nn
\ea

\subsubsection{Covariant gauge}

If we follow the literature \cite{Hinfa,HinfRG,Barbon:2009ya} and specialize to unitary gauge at this point, $\H^\ssT = (0, h)/\sqrt2$, the kinetic terms involve only a single scalar and so one can define a new field, $h = h(\chi)$, such that $\chi$ has a canonical kinetic term, $\frac12 \, \partial \chi \partial \chi$. The same {\em cannot} be done in a general gauge because in this case the scalar target-space metric is not flat. To see this, we write the doublet in terms of four real fields, with $\sqrt{2} \, \H^\ssT/M_p =(\phi_1, \phi_2,\phi_3, \phi_4)$. Then the kinetic sector becomes $\frac12 M_p^2 \, g^{\mu\nu} \G_{ij}(\phi) \,
\partial_{\mu} \phi^i \, \partial_{\nu} \phi^j$, and so the target-space metric is read off as
\ba
    \G_{ij}(\phi) = f \, \delta_{ij} + 6
 \,  \xi^2 \, f^2 \phi_i \, \phi_j \,,
\ea
where $f^{-1} = (1 +  \xi \, \vec \phi^2)$, with $\vec \phi^2 = \sum_i \phi^2_i$. It is straightforward to compute the Riemann tensor, ${\R^i}_{jkl}$, for this metric, which does not vanish. Since the target-space metric is not flat, there does not exist a field redefinition which everywhere sets $\G_{ij} = \delta_{ij}$. At best one can do so only at a specific point in field space, $\phi^i = \phi^i_0$ (or along a geodesic).

The significance of this observation is that the matrix element of the derivative interaction appearing in the second line of eq.~\pref{dim6} survives field redefinitions to give a direct contribution to $\H \, \H \rightarrow \H \, \H$ scattering. The large-$E$ limit of the resulting amplitude is given by
\be
 \A(E) \simeq (E/\Lambda)^2,
\ee
which agrees with the Jordan-frame result.

\subsubsection{Unitary Gauge}

What happens if one insists on using unitary gauge, so that the kinetic sector can be canonically normalized? In this case the required redefinition of the Higgs field is $h \to \chi$ where
\ba
 \frac{d \, \chi}{d \, h} =  \frac{\left[1+
 (\xi +  6 \, \xi^2) \,(h/M_p)^2\right]^{1/2}}{1 + \xi \,
 (h/M_p)^2}.
\ea
As was shown in \cite{Barbon:2009ya} in the small-field limit this implies $h$ and $\chi$ are related by
\ba
 h = \chi \left[1 - \chi^2/\Lambda^2 \right] + \cdots \,,
\ea
and so $\langle \chi \rangle = v_\ssH(1 + v_\ssH^2/\Lambda^2) + \mathcal{O}(v_\ssH^5/\Lambda^4)$. This introduces $\Lambda$ into the scalar potential, such as by converting
\ba
 \lambda_\ssH h^4 = \lambda_\ssH \chi^4
 - \frac{4\lambda_\ssH}{\Lambda^2} \, \chi^6 + \cdots,
\ea
and so on. Of course such terms do not change particle physics applications, where generally one stops at quartic order in $\chi$ because $\chi \sim v_\ssH \ll \Lambda$. Ref.~\cite{Barbon:2009ya}'s point is that such terms become
important for large-field applications -- such as inflation -- and they in particular raise the issue of what justifies keeping only a quartic potential in $\H$ (or quadratic function pre-multiplying $R$) in the first place -- as is crucially required to obtain a flat inflationary potential.

If the non-minimal kinetic terms are not present in unitary gauge, where does the large-$E$ behaviour of the $\H \, \H \rightarrow \H \, \H$ scattering amplitude come from? The key point is that the transformation to unitary gauge moves the three would-be Goldstone modes into the longitudinal components of the gauge bosons, suggesting we should seek the large-$E$ unitarity problems from graphs involving these eaten degrees of freedom.

For instance, consider the contribution to $\chi$ scattering from longitudinal $Z$ bosons coming from the $ \mathcal{O}(1/\Lambda^2)$ corrections to the SM Higgs/gauge-boson interactions (where $\tilde{h} = h -  \langle h \rangle$ and $\tilde \chi = \chi - \langle \chi \rangle$),
\ba
 && - \frac{M_\ssZ^2}{2} Z_\mu Z^\mu
 \left(\frac{ 2\tilde h}{v_\ssH}
 +  \frac{\tilde{h}^2}{v_\ssH^2}\right) =
 - \frac{M_\ssZ^2}{2 \, v_\ssH^2} Z_\mu Z^\mu
 \tilde \chi^2 \left[1 - \frac{12 \, v_\ssH^2}{\Lambda^2}
 \right] \nn \\
 && \qquad\qquad\qquad - \frac{M_\ssZ^2}{v_\ssH} Z_\mu Z^\mu
 \tilde \chi \left[1 - \frac{3 \,  v_\ssH^2}{\Lambda^2}
 \right] + \cdots \,,
\ea
The amplitude for $\chi Z_\ssL\to \chi Z_\ssL$ receives a $1/\Lambda^2$ correction of order
\ba
 A(E) &\simeq& \frac{M_\ssZ^2}{\Lambda^2}
 \, \eps_{\ssL\,\mu} \, {\eps_\ssL}^\mu, \nn \\
 &\simeq& (E / \Lambda)^2 \,,
\ea
which uses the fact that the longitudinal polarization vectors behave as ${\eps_\ssL}^\mu \sim p^\mu/M_\ssZ$ in the
large-momentum limit, and that this unitarity violation famously cancels \cite{CLT} for the Standard Model in the $\Lambda \to \infty$ limit.

Similar results hold for other scattering amplitudes involving these degrees of freedom. For example in  $W_\ssL \, W_\ssL \to W_\ssL \, W_\ssL$ scattering the unitarity violation comes about from the $ \mathcal{O}(1/\Lambda^2)$ correction due to the non-minimal coupling in the higgs exchange graphs in $W_\ssL \, W_\ssL \to W_\ssL \, W_\ssL$ while the other graphs involving the operators with couplings to the $Z, \gamma$ and the four point $W_\ssL \, W_\ssL W_\ssL \, W_\ssL$ operator are unaffected by the non-minimal coupling. This interferes with the cancelation of the
corresponding contributions to $W_\ssL \, W_\ssL \to W_\ssL \, W_\ssL$ scattering in the SM \cite{CLT}, inducing the same energy dependence of the amplitude $A(E) \simeq (E / \Lambda)^2.$

\section{Conclusions}

All roads lead to Rome: Higgs scattering amplitudes computed in the Einstein frame and in different gauges have the same unitarity problems at energies of order $\Lambda = M_p/\xi$ as do explicit Jordan frame calculations \cite{HW,Atkins:2010eq}. This is consistent with the general power-counting arguments given in ref.~\cite{Burgess:2009ea}, but contrary to the recent claims of \cite{Lerner:2009na}. Unitarity problems like these indicate a failure of the semiclassical approximation, such as those underlying an inflationary analysis. Inflation could nonetheless occur, but to the extent that it does so at scales at or above $\Lambda$ its justification is better made using whatever physics
intervenes at this scale to resolve the unitarity problem.

\section{Addendum}

Shortly after this paper was posted ref.~\cite{Hertzberg:2010dc} appeared, as did an addendum to ref.~\cite{Lerner:2009na}, both commenting on our previous work, \cite{Burgess:2009ea}, as well as this paper. With this Addendum we briefly offer final comments on each of these subsequent developments.

Both \cite{Hertzberg:2010dc}, and now \cite{Lerner:2009na}, agree that unitarity issues arise at the scale $\Lambda$ if the inflaton is part of a Higgs doublet, confirming that the earlier power counting results \cite{Burgess:2009ea} correctly identified the scale where the low-energy approximation fails. Ref.~\cite{Hertzberg:2010dc} also confirms part of the story told in this paper as to how unitarity problems arise in detail in the Einstein frame: through non-minimal kinetic interactions amongst
the scalars in a covariant gauge.

However ref.~\cite{Hertzberg:2010dc} points out, correctly, that these same graphs do not pose a tree-level unitarity problem for $\phi \, \phi \to \phi \, \phi$ scattering\footnote{We use $\phi$ rather than $h$ to denote the non-canonical Einstein frame scalar to emphasize that it is a singlet field, and not the Higgs boson.}
in the simpler model where the scalar is a real singlet, involving no would-be Goldstone modes (a similar point is made in v2 of ref.~\cite{Lerner:2009na}). This is clearest in the Einstein frame, since in this case there is no obstacle to canonically normalizing the fields. Then the only $\Lambda$-dependence in the quartic interactions of the scalar potential are: $\delta V \simeq m_\phi^2 \, \chi^4/ \Lambda^2$, leading to a 4-point tree-level $\phi$-scattering amplitude that does not grow like a power of energy. As always, it should be possible to arrive at this same
conclusion in all frames, and for the Jordan frame
ref.~\cite{Hertzberg:2010dc} makes the point that this does not contradict power-counting arguments (or the calculations of \cite{HW}, who compute only partial-wave amplitudes), because although each individual graph (or partial wave) grows quadratically with energy, the leading behaviour when summed over all channels is proportional to $(s + t + u)/\Lambda^2 \propto m_\phi^2/\Lambda^2$ \cite{Huggins:1987ea}. The same absence of growing contributions can also be seen in the Einstein frame
without canonically normalizing the scalar since the dangerous kinetic interaction, ${\cal L} \simeq \phi^2
\partial_\mu \phi \, \partial^\mu \phi/\Lambda^2$ can be
rewritten as a total derivative plus a term ${\cal L} \simeq
\phi^3 \Box \phi/\Lambda^2$, that becomes ${\cal L} \simeq
m_\phi^2 \, \phi^4/\Lambda^2$ using the lowest-order equations of motion (for a detailed justification of this last step, see for example \cite{GREFT}).

Based on these results ref.~\cite{Hertzberg:2010dc} goes on to argue that unitarity problems are unlikely to arise at scale $\Lambda$ for singlet scalar models, and also for the $R^2$ inflation model, which can be rewritten as a singlet-scalar model and was criticized on these grounds in ref.~\cite{Burgess:2009ea}.

Although we agree with the discussion of tree-level $\phi \, \phi \to \phi \, \phi$ scattering, we disagree with the conclusion that this suffices to establish unitarity (for singlet scalars or for $R^2$ inflation) at the scale $\Lambda$. In particular, power counting also indicates that processes like $\phi \, \phi \to \phi \, \phi \, \phi \, \phi$ have tree-level cross sections that behave as $\sigma \simeq E^2/\Lambda^4$, such as would arise in Einstein frame with canonically normalized fields from scalar potential interactions of the form $\delta V \simeq \lambda \,
\chi^6/\Lambda^2$. Again, for these processes unitarity implies $E < \Lambda$. The same is true for scattering involving more $\phi$ particles, and/or involving higher loops (although if higher loops are at work the scale of unitarity violation is slightly raised to $\sim 4\pi \Lambda$ because the accompanying loop factors of $1/16\pi^2$).

It is logically possible that cancelations amongst graphs
conspire to suppress contributions to scattering relative to
power-counting estimates, order by order in the loop expansion. This is what would be required to allow the theory to make sense at energies above the scale $\Lambda$ (despite power-counting indications to the contrary). The only known theories where this happens are those where the cancelations are enforced by a (possibly approximate) symmetry, and the great interest of these scalar inflation models makes it worthwhile to explore this possibility in more detail.

In the absence of any evidence of this form  we stand by the
conclusions of ref.~\cite{Burgess:2009ea}. If anything, the
evidence is that the cancelation in $\phi \, \phi \to \phi \, \phi$ at tree level is an accident of the simplicity of this process, since the same arguments that indicate that interactions like $\phi^2 (\partial_\mu \phi \, \partial^\mu \phi)$ are not problematic, do not apply to more general Einstein-frame interactions, like $\phi^2 (\partial_\mu \phi \, \partial^\mu \phi)^2$ or $f(\phi) \,\partial_\mu \phi \, \partial^\mu \phi$. Even if not present at the classical level, such interactions are inevitably generated by loops. Of course, the burden of proof lies on any proponent of a particular inflationary model to demonstrate control over the approximations made, but we are encouraged that attempts along these lines are now starting to be done.

\subsection{Note Added on Background Dependence}

Although somewhat tangential to this paper's main line of development, we comment here on the sensitivity of the power-counting arguments used in this paper (and \cite{Burgess:2009ea}), to the various background fields present.

There are two kinds of background fields to discuss for Higgs inflation models, the curvature of the background metric and the value of the Higgs field itself. Zero curvature and $\langle h \rangle \ll M_p$ is the regime for connecting to potential Higgs physics at the LHC, while $R \simeq H^2$ and $\sqrt{\xi} \, \langle h \rangle \gg M_p$ is the putative domain of inflation. Both are clearly required to relate quantities determined in colliders to quantities measured from CMB data, as advocated in \cite{HinfRG}, when using RG methods to run the effective theories from one scale to another. One might worry (as does footnote 1 in v3 of \cite{Germani:2010ux}) whether the constraints described above (or in \cite{Burgess:2009ea, Barbon:2009ya}) are restricted only to the first regime of negligible background fields.

The inflationary background fields do not provide a loophole for our arguments for the following reasons. First, it is a misconception that the power-counting arguments of \cite{Burgess:2009ea} are limited to an expansion about flat space with vanishing Higgs field. Even though they are often derived (as in \cite{GREFT})  in momentum space, these arguments are based on dimensional analysis. As such, all they require is that the low-energy physics be characterized by a single scale, say $H$, in order to properly keep track of the dependence of observables on powers like $H/\Lambda$. They therefore apply equally well if computed in position space using the full de Sitter propagators. 

A full treatment of the implications of limit $\sqrt{\xi} \, \langle h \rangle \gg M_p$ is necessarily more thorny, largely due to the lack of control (emphasized in \cite{Barbon:2009ya}) over the form of the lagrangian in this limit. Assuming the potential and Einstein terms to be precisely quartic and quadratic polynomials in this limit in the Jordan frame, (the potential is more complicated in the Einstein frame, and the necessity to choose a frame when specifying the functional form is unlikely to be ensured by a physical condition such as a symmetry), one might wonder if having $\sqrt{\xi} \, \langle h \rangle \gg M_p$ in front of $R$ might change the Jordan-frame power-counting in this limit. 
As discussed in \cite{Germani:2010ux,Germani:2010gm} this condition can lead to the potentially confusing conclusion that the
cut off scale in the Jordan frame is different than the cut off scale in the Einstein frame when a large Higgs vev is present. 
This would appear to contradict the general result that ratios of physical mass scales (like $\Lambda/M_p$) must be frame independent.

However, an explicit check in \cite{Germani:2010gm} shows that one still finds $H \propto \Lambda$ up to the same factors of the Higgs self-coupling $\lambda$ found in \cite{Burgess:2009ea}. Note that the apparent cut off scale  difference is despite the fact that the arguments of \cite{Burgess:2009ea} apply equally well in either frame, since they do not require the fluctuations fields to have been canonically normalized. This apparent contradiction is resolved with the use a physical definition of  $M_p$  when comparing the Einstein and Jordan frame results when  $\xi \langle h^2 \rangle  \gg M_p^2$. Using a physical definition of  $M_p$
(such as the strength of gravity between two test masses) reveals that $M_{p}$ also changes in going between frames, and in terms of a physical $M_p^{phys}$ the same cut off scale 
is obtained irrespective of the Higgs vev, ie $\Lambda \simeq M_p^{phys}/\xi$

\section*{Acknowledgements}

We would like to acknowledge a helpful correspondence with
R.~N.~Lerner and J.~McDonald, and thank J.L.F. Barb\'on for useful
discussions. M.T thanks S. Shandera for a kind code donation to
the effort. This work was partially supported by funds from the
Natural Sciences and Engineering Research Council (NSERC) of
Canada. Research at the Perimeter Institute is supported by the
Government of Canada through Industry Canada and by the Province
of Ontario through the Ministry of Research \& Innovation.


\newpage

\begin{thebibliography}{99}

\bibitem{earlierNMC}
 D.~S.~Salopek, J.~R.~Bond and J.~M.~Bardeen,
  Phys.\ Rev.\  D {\bf 40}, 1753 (1989);

   D.~La and P.~J.~Steinhardt,
  Phys.\ Rev.\ Lett.\  {\bf 62} (1989) 376
  [Erratum-ibid.\  {\bf 62} (1989) 1066].

  R.~Fakir and W.~G.~Unruh,
  Phys.\ Rev.\  D {\bf 41}, 1783 (1990);

 D.~I.~Kaiser,
  Phys.\ Rev.\  D {\bf 52}, 4295 (1995)
  [arXiv:astro-ph/9408044];

    E.~Komatsu and T.~Futamase,
  Phys.\ Rev.\  D {\bf 59}, 064029 (1999)
  [arXiv:astro-ph/9901127].

\bibitem{Hinfa}
  F.~L.~Bezrukov and M.~Shaposhnikov,
  Phys.\ Lett.\  B {\bf 659}, 703 (2008)
  [arXiv:0710.3755 [hep-th]].

  \bibitem{HinfRG}
 Andrea De Simone, Mark P. Hertzberg, and Frank Wilczek,
 [arXiv:0812.4946];

 F.L. Bezrukov and M. Shaposhnikov,
 [arXiv:0812.4950]. 

\bibitem{WMAP}
 G. Hinshaw {it et.al.}
 ``Five-Year Wilkinson Microwave Anisotropy
 Probe (WMAP) Observations:Data Processing, Sky Maps, \& Basic Results",
 [arXiv:0803.0732].

\bibitem{Burgess:2009ea}
  C.~P.~Burgess, H.~M.~Lee and M.~Trott,
  JHEP {\bf 0909}, 103 (2009)
  [arXiv:0902.4465 [hep-ph]].

\bibitem{GREFT}
  J.~F.~Donoghue,
  [arXiv:gr-qc/9512024];

  C.~P.~Burgess,
  Living Rev.\ Rel.\  {\bf 7} (2004) 5
  [arXiv:gr-qc/0311082].

\bibitem{HW}
   T.~Han and S.~Willenbrock,
  Phys.\ Lett.\  B {\bf 616} (2005) 215
  [arXiv:hep-ph/0404182].

\bibitem{Bezrukov:2009db}
  F.~Bezrukov and M.~Shaposhnikov,
  JHEP {\bf 0907}, 089 (2009)
  [arXiv:0904.1537 [hep-ph]].

\bibitem{Barbon:2009ya}
  J.~L.~F.~Barbon and J.~R.~Espinosa,
  Phys.\ Rev.\  D {\bf 79}, 081302 (2009)
  [arXiv:0903.0355 [hep-ph]].

\bibitem{BE}
 C.P.~Burgess V. Di Clemente and J.R. Espinosa,
 JHEP 0201 (2002) 041 [hep-ph/0201160];

\bibitem{Lerner:2009na}
  R.~N.~Lerner and J.~McDonald,
  arXiv:0912.5463.

\bibitem{CLT}
 J.M.~Cornwall, D.N. Levin and G.~Tiktopoulos, Phys.\ Rev.\ {\bf
 D10} (1974) 1145;

  B.W.~Lee, C.~Quigg and H.B.~Thacker,
  Phys.\ Rev.\  D {\bf 16}, 1519 (1977).

\bibitem{Atkins:2010eq}
  M.~Atkins and X.~Calmet,
  arXiv:1002.0003 [hep-th].

\bibitem{Hertzberg:2010dc}
  M.~P.~Hertzberg,
  arXiv:1002.2995 [hep-ph].

\bibitem{Huggins:1987ea}
  S.~R.~Huggins and D.~J.~Toms,
  Class.\ Quant.\ Grav.\  {\bf 4}, 1509 (1987).

\bibitem{Germani:2010ux}
  C.~Germani and A.~Kehagias,
  JCAP {\bf 1005} (2010) 019
  [arXiv:1003.4285 [astro-ph.CO]].

\bibitem{Germani:2010gm}
  C.~Germani and A.~Kehagias,
  arXiv:1003.2635 [hep-ph].


\end{thebibliography}
\end{document}